\documentclass[%
 reprint,
superscriptaddress,
 amsmath,amssymb,
 aps,
prb,
]{revtex4-2}

\usepackage{graphicx}%
\usepackage[linktocpage,colorlinks,breaklinks,linkcolor=blue,citecolor=blue,urlcolor=blue]{hyperref}%

\begin{document}

\title{Quantum Monte Carlo study of Doppler broadening of positron annihilation radiation in semiconductors and insulators}%

\author{K. A. Simula}
\affiliation{Department of Physics, P.O. Box 43, FI-00014 University of Helsinki, Finland}
\author{J. Härkönen}
\affiliation{Department of Physics, P.O. Box 43, FI-00014 University of Helsinki, Finland}
\author{I. Zhelezova}
\affiliation{Department of Physics, P.O. Box 43, FI-00014 University of Helsinki, Finland}
\author{N. D. Drummond}
\affiliation{
Department of Physics, Lancaster University, Lancaster LA1 4YB, United Kingdom 
}%
\author{F. Tuomisto}
\affiliation{Department of Physics, P.O. Box 43, FI-00014 University of Helsinki, Finland}
\author{I. Makkonen}%
\affiliation{Department of Physics, P.O. Box 43, FI-00014 University of Helsinki, Finland}

\date{\today}%

\begin{abstract}
  Positron annihilation in solid state matter can be utilized to detect and identify open-volume defects. The momentum distribution of the annihilation radiation is an important observable in positron-based measurements, and can reveal information on the chemical surroundings of the defect sites. In this work we present a variational quantum Monte Carlo method for simulation of the momentum densities of annihilating electron-positron pairs in semiconductors and insulators. We study finite-size effects, effects of lattice vibrations, and different levels of trial wave functions. Small simulation cells and simple wave function forms are found to be sufficient for accurate calculations in simulation of pristine lattices, enabling cheap accumulation of results. We compare calculated predictions of the Doppler broadening of the 511-keV $2\gamma$ annihilation line of the in aluminium nitride and silicon against experimental data measured from reference samples. Our results achieve better agreement with experiments in the these materials than conventional state-of-the-art methods, and proves that direct modeling of the electron-positron correlations is important for a supporting theory of positron annihilation sprectroscopies.

\end{abstract}

\maketitle

\section{Introduction}

Positron annihilation spectroscopy is a powerful method for detecting, quantifying, and identifying defects in crystalline solids such as metals, alloys, semiconductors, and complex oxides \cite{TuomistoRMP2013}. In order to determine the atomic and electronic structures responsible for the observed annihilation radiation, theoretical considerations and computational methods are needed. Accurate calculations require well defined and practical \textit{ab initio} methods. 

So far, the only practical method to support positron experiments for defects is the two-component density-functional theory (DFT) for electron-positron systems \cite{BoronskiPRB1986}. It is a powerful method for modeling solid-state systems with positrons annihilating either from a delocalized state or from trapped states in open-volume defects.
At best, a combination of high-quality experiments and modeling can help to identify point defects in semiconductors \cite{WiktorPRB2016,TuomistoPRL2017}, complex oxides \cite{KarjalainenPRB2020,MakkonenJPCM2016} as well as semiconductor alloys \cite{ishibashi2016,ProzheevaAPL2017,ishibashi2019} and novel metallic alloys \cite{TuomistoAM2020}, to name only a few recent examples. 
However, in addition to the usual electronic exchange-correlation energy, one has to approximate and parametrize the electron-positron correlation energy and the so-called enhancement factor that describes the screening of the positron by electrons when calculating the positron annihilation rate. Momentum-space quantities such as the annihilating-pair momentum density (APMD), which is directly related to the Doppler broadening of the annihilation radiation, are not  well-defined in the DFT framework \cite{MakkonenPRB2014}. Instead, approximations are needed, which again are based on parametrized enhancement factors (see, for example, Refs.~\cite{Daniuk1987,*Jarlborg1987} and \cite{alatalo1996}). Often the only way to justify the choice of a particular combination of approximations for a given problem is experimental validation \cite{MakkonenPRB2006}.

Modeling of correlated densities or momentum space quantities from first principles without the aforementioned approximations requires in practice the application of many-body wave function methods \cite{DrummondPRL2011} or many-body perturbation theory \cite{green2015,hofierka2022}. Thanks to the development of the available computing capacity and suitable methods this is becoming possible. Recently it has been demonstrated for a small benchmark set of systems that quantum Monte Carlo (QMC) calculations based on variational and diffusion Monte Carlo (VMC and DMC) can be used for real inhomogenous periodic solids \cite{SimulaPRL2022}. The QMC method can predict parameter-free positron lifetimes of defect-free solids with an accuracy that is on average better than that of existing two-component DFT approximations \cite{SimulaPRL2022}.

In this work, we extend the QMC simulation of positrons to estimation of APMDs for the same insulators and semiconductors as in Ref.~\cite{SimulaPRL2022}: C and Si in the diamond structure, as well as wurtzite AlN. This is a set that has been earlier used in successful simulation of positron lifetimes with QMC~\cite{SimulaPRL2022}, contains one insulator and two semiconductors, most of which are characterized accurately enough experimentally using both positron lifetime and Doppler broadening. Also, they do not have electrons in the $d$ shell, whose contribution to annihilation is usually significant. Most pseudopotentials do not explicitly consider the $d$ electrons but include them in the frozen core. Using VMC, we focus on experimental validation of the method's applicability in the prediction of the Doppler broadening of the annihilation radiation in AlN and Si. Measurement of the Doppler broadening allows for detailed identification of the defects thanks to the sensitivity to the electron configurations in the immediate surroundings of the positron localization site. Accurate interpretations require strong theory support. Even though we expect the method to work for metals, we exclude them here due to the larger computational cost associated with the increased momentum resolution needed to describe Fermi surface signatures correctly.

By establishing a workflow for VMC simulation of APMDs, we study the required accuracy for the wave functions and treatment of finite-size effects. Also, vibrational effects on Doppler spectra are studied and found to be insignificant. With highly optimized wave functions and suitable simulation cell sizes,  we calculate predictions for the Doppler broadening in Si and AlN, and compare the results against experiments. The comparison shows that our method outperforms DFT-based methods.

Together with the recent milestone of using QMC to accurately predict positron lifetimes in solids \cite{SimulaPRL2022}, this work generalizes the QMC method as a practical tool to support positron-based spectroscopies. This work together with Ref.~\cite{SimulaPRL2022} provides a consistent picture of the accuracy of the method when both positron lifetime and Doppler broadening are considered.  Of particular importance is the finding that, unlike with lifetime calculations, in calculating the Doppler broadening small simulation cell sizes are convergent and the backflow corrections are not needed. Hence accurate Doppler broadening results can be achieved with small computational cost, as opposed to lifetime calculations, where at least the backflow function is needed \cite{SimulaPRL2022}. 

Promising future applications include simulations of positrons trapped at large void-like defects or the study of surface states, where non-local electron-positron correlations dominate and local or semilocal correlation functionals of the two-component DFT fail \cite{shi2018,yang2022}. These extensions to the applicability of QMC enable it  to support positron spectroscopic methods detecting and identifying open-volume defects in crystalline matter \cite{TuomistoRMP2013} or surface-layer sensitive surface techniques such as positron-annihilation induced Auger spectroscopy \cite{chirayath2017,*fairchild2022}. 

\section{Theory and computations}

In positron annihilation experiments on solids, positrons are implanted into the lattice one at a time. They thermalize rapidly and often get trapped in open-volume defects before annihilating with an electron. The technique probes the ground state of the electron-positron system. The APMD is reflected in the annihilation $\gamma$ radiation. The Doppler broadening of the 511-keV annihilation line measures essentially the longitudinal component of the momentum density of annihilating pairs.

We simulate the ground state of inhomogenous crystal lattices with a single thermalized positron using periodic boundary conditions. We use the VMC method \cite{foulkes2001} as implemented in the \textsc{casino} code \cite{needs2009,needs2020}, with Slater-Jastrow (SJ) and Slater-Jastrow-backflow (SJB) wave functions \cite{jastrow1955,rios2006}:
\begin{align}
\label{equation:slater-jastrow wf, qualitative}
\Psi_{\text{SJ}}(\mathbf{R}) &= e^{J(\mathbf{R})}\left[ \phi^l%
(\mathbf{r}_{i\uparrow}) \right]\left[ \phi^m%
(\mathbf{r}_{j\downarrow}) \right]\phi_+(\mathbf{r}_+),\\
\Psi_{\text{SJB}}(\mathbf{R}) &= e^{J(\mathbf{R})}\left[ \phi^l%
\boldsymbol{(}\mathbf{r}_{i\uparrow} -\boldsymbol{\xi}_{i\uparrow}(\mathbf{R})\boldsymbol{)} \right]\left[ \phi^m%
\boldsymbol{(}\mathbf{r}_{j\downarrow}-\boldsymbol{\xi}_{j\downarrow}(\mathbf{R})\boldsymbol{)} \right]\nonumber\\ 
&\times \phi_+\boldsymbol{(}\mathbf{r}_+-\boldsymbol{\xi}_+(\mathbf{R})\boldsymbol{)}, \nonumber
\end{align}
where $\mathbf{R}=(\mathbf{r}_+,\mathbf{r}_1,\ldots,\mathbf{r}_N)$ is a vector of length $3(N+1)$ containing the coordinates of the positron and  $N$ electrons, respectively. The SJ wave functions $\Psi_{\text{SJ}}$ consist of Slater determinants, denoted by $[...]$-brackets in Eq.~(\ref{equation:slater-jastrow wf, qualitative}), and constructed from single-particle Kohn-Sham orbitals $\phi$~\cite{kohn1965}. The determinants are multiplied by a Jastrow factor $\exp({J(\mathbf{R})})$ describing the interparticle correlations. The Jastrow factor consists of parametrized polynomial components describing two-particle, particle-nucleus and three-particle correlations~\cite{drummond2004}. In the SJB wave functions, the particle coordinates in the orbitals are replaced by quasiparticle coordinates $\mathbf{r}-\boldsymbol{\xi}(\mathbf{R})$, in which $\boldsymbol{\xi}(\mathbf{R})$ is represented by a parametrized polynomial of interparticle distances.

The single-particle electronic orbitals $\{\phi\}$ are obtained from a DFT simulation using Quantum \textsc{espresso} \cite{giannozzi2020} and the PBE functional \cite{PerdewPRL1996}. To obtain the positron orbital we use our own positron package \cite{torsti2006}. 

Before accumulating statistics on momentum densities, we first optimize the Jastrow factor with variance minimization \cite{drummond2004}, followed by optimization of both the Jastrow factor and backflow transformations by minimizing the energy \cite{umrigar2007}. The parametrizations are consistent with the positron lifetime studies performed in Ref.~\cite{SimulaPRL2022}, which also describes in detail how the Slater part of the wave functions is constructed. 

Twisted boundary conditions \cite{lin2001} are imposed so that the wave function obtains a phase under translation of an electron through a simulation cell lattice vector $\mathbf{R}_s$,
\begin{equation}
\Psi_{\mathbf{k}_s}(\mathbf{r}_+,\ldots,\mathbf{r}_i+\mathbf{R}_s,\ldots,\mathbf{r}_N)=\exp(i\mathbf{k}_s\cdot \mathbf{R}_s)\Psi_{\mathbf{k}_s}(\mathbf{r}_+,\ldots,\mathbf{r}_N).
\end{equation}

Twist offsets $\mathbf{k}_s$ are set in a regular grid in the irreducible wedge of the $1$st Brillouin zone of the simulation cells. Regarding translations of the positron, we do not use twisted boundary conditions, since in the usual experimental case we have only one delocalized positron in a crystal lattice at a time, which corresponds physically to a positron orbital with $\mathbf{k}=\mathbf{0}$.

Multiple finite-size effects, originating from the finite simulation cell and discrete momentum grid of the wave function, can bias computed energies and positron observables.  Kinetic energy bias due to discrete momentum grid \cite{holzmann2016} and the effects from neglecting long-range correlation effects can be reduced with larger simulation cell sizes and twist averaging. Twist averaging is also used to obtain a sufficient momentum resolution in a momentum density calculation. Earlier studies \cite{SimulaPRL2022} of positron relaxation energies and positron-electron contact pair correlation functions show that $16$-atom face-centered cubic simulation cells for C and Si and $16$-atom hexagonal cells for AlN with twist grids of $4\times 4\times 4$ and $4\times 4\times 5$, respectively, are already fairly convergent. However, we have done further studies on the finite-size effects in Doppler spectra and present the results below.

In the case of 2$\gamma$ annihilation (spin-singlet electron-positron pairs) the APMD corresponds to the net momentum density of the 511-keV annihilation photons. The momentum density for a wave function $\Psi_{\mathbf{k}_s}$ is a set of values $\rho(\mathbf{p}_i)$ on a momentum grid, where $\mathbf{p}_i=\mathbf{k}_s+\mathbf{G}_i$. Here, $\mathbf{G}_i$ are the reciprocal lattice points of the simulation cell. $\rho(\mathbf{p}_i)$ can be computed with
\begin{align}
  \label{equation: apmd formula for vmc}
\rho(\mathbf{p}_i)=&\pi r_0^2c \int d\mathbf{R}\,\Psi_{\mathbf{k}_s}^*(\mathbf{r}_1,\mathbf{r}_1,\ldots,\mathbf{r}_N)\\
\times & e^{i\mathbf{p}_i\cdot(\mathbf{r}_+-\mathbf{r}_1)}\hat{O}^{s}_{i}\Psi_{\mathbf{k}_s}(\mathbf{r}_+,\mathbf{r}_+,\ldots,\mathbf{r}_N),\nonumber
\end{align}
where $r_0$ is the classical radius of electron and $c$ is the speed of light \textit{in vacuo} \cite{ryzhikh1999}. $\hat{O}^{s}_{i}$ is the spin-projection operator to the singlet-state of the electron-positron pair.
In VMC, we propagate the $N+1$ particles using the Metropolis algorithm and sample the APMD using 
\begin{widetext}
  \begin{align}
    \label{equation: doppler projection}
\rho(\mathbf{p}_i)=& \left\langle \frac{\Psi_{\mathbf{k}_s}^*(\mathbf{r}_1,\mathbf{r}_1,\ldots,\mathbf{r}_N)\hat{O}^{s}_{i}\Psi_{\mathbf{k}_s}(\mathbf{r}_+,\mathbf{r}_+,\ldots,\mathbf{r}_N)}{|\Psi_{\mathbf{k}_s}(\mathbf{r}_+,\mathbf{r}_1,\ldots,\mathbf{r}_N)|^2}e^{i\mathbf{p}_i\cdot(\mathbf{r}_1-\mathbf{r}_+)}\right\rangle_{|\Psi_{\mathbf{k}_s}|^2}.
\end{align}
\end{widetext}
The averaging involves in practice also a permutation over opposite spin electron indices, which does not affect the expectation value due to exchange antisymmetry, but reduces the standard error in the mean for a given number of VMC-sampled configurations.

The agreement between our SJ- and SJB-VMC results for the Doppler projections, estimated with Eq. (\ref{equation: doppler projection}) from the APMD data, are converged with respect to the variational freedom in the wave functions.

\begin{figure*}
  \includegraphics[scale=.8]{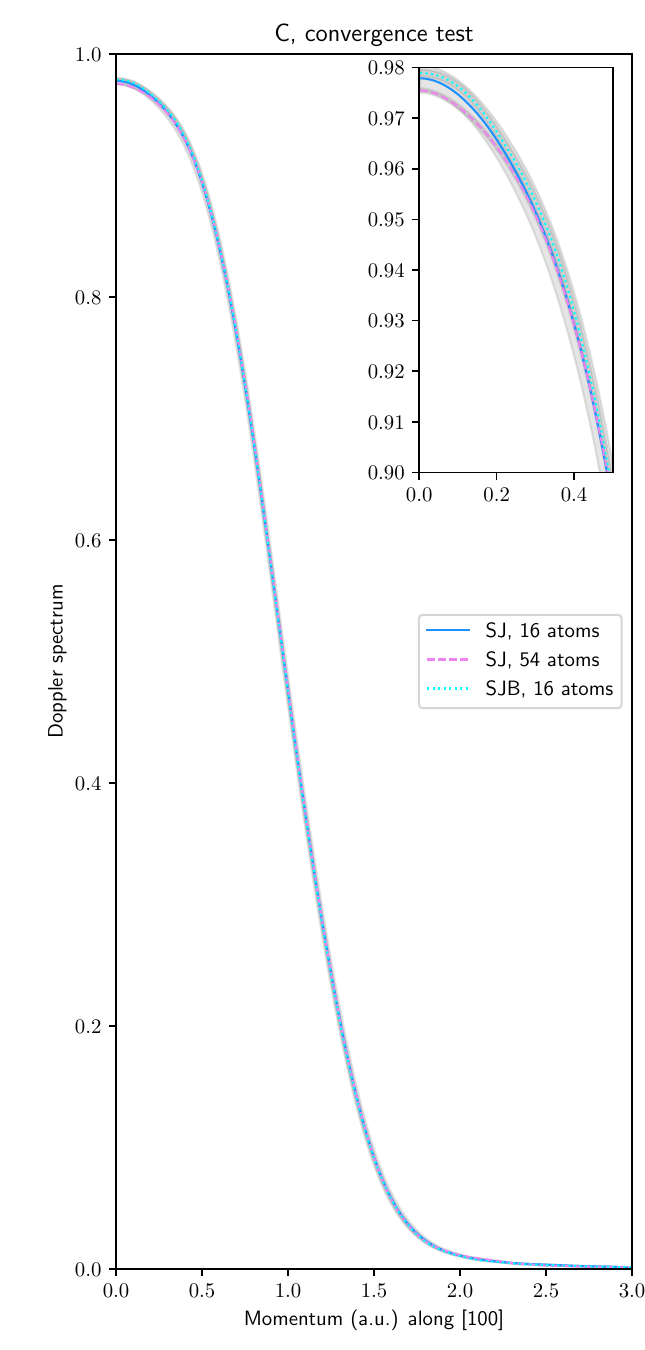}
  \includegraphics[scale=.8]{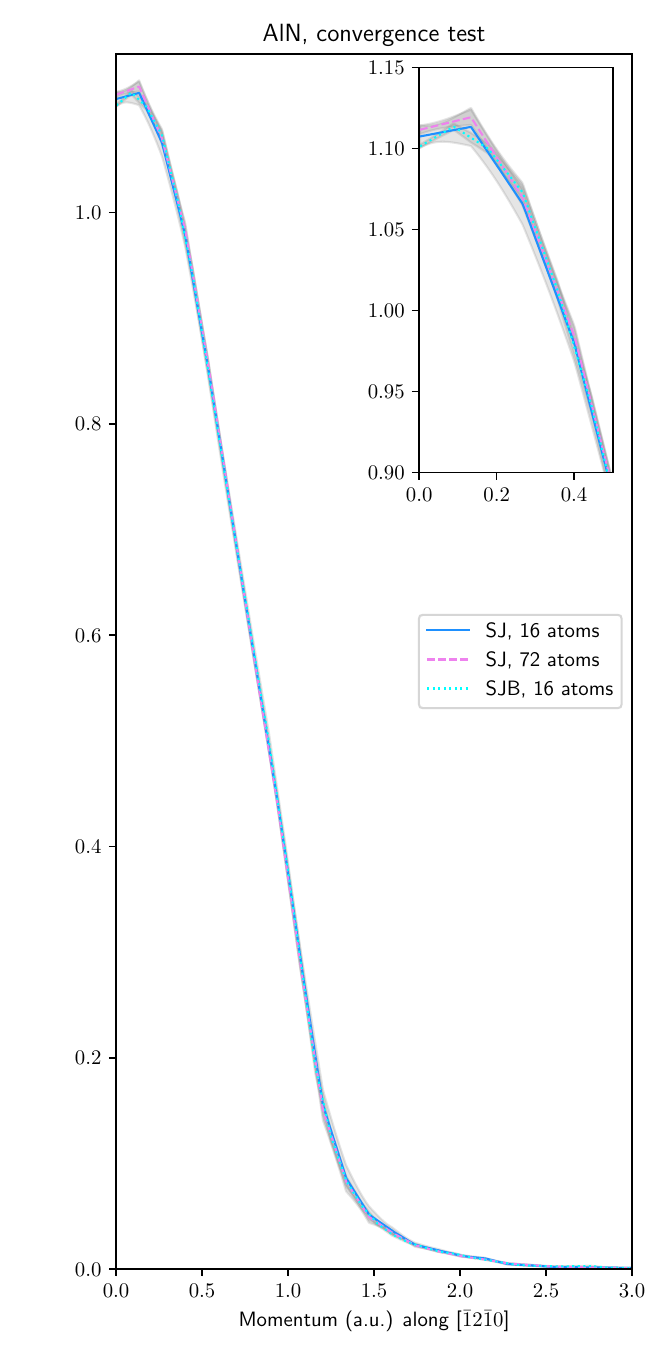}
  \caption{\label{figure: c convergence} Unconvoluted Doppler spectra of C (left) and AlN (right) in the [100]- and [$\bar{1}2\bar{1}0$]-directions, respectively. In both materials, the APMD is computed with two different simulation cell sizes (blue: $16$-atom cell, violet: $54$-atom cell in C and $72$-atom cell in AlN) with a SJ wave function and with a SJB wave function in the smaller simulation cell (light blue). }
\end{figure*}

The grid of momentum density values is limited by a suitable cutoff value, so that $|\mathbf{p}_i|<p_{cut}$. In the VMC simulations, we chose $p_{cut}=7$ a.u., but found the Doppler spectra to be convergent with values as low as $\sim 5$ a.u. As mentioned, twist averaging is used to increase the momentum resolution of our APMD estimates. This means that with $N_s$ symmetrically inequivalent twists we must construct $N_s$ wave functions and estimate the APMD for each in an independent simulation. However, we have optimized the Jastrow factor only at the $\Gamma$-point twist. 

We calculate 1D Doppler projections from the 3D APMDs as
\begin{equation}
  \label{equation: doppler projection formula}
  \rho(p_L)=\int \int dp_x dp_y \rho(\mathbf{p}),
\end{equation}
where $p_L$ is the longitudinal momentum component of the annihilation photons. The projection is made over the orthogonal plane, by first filling the momentum grid by non-intersecting tetrahedra with a Delaunay tetrahedralization algorithm \cite{shewchuk1999}. Then, the projected $\rho(p_L)$ values are computed with a linear interpolation algorithm \cite{matsumoto2004}. The analysis of statistical error is based on treating the twists $\mathbf{k}_s$ as independent measurements. Before comparing with experiment, the computed Doppler spectra are normalized to unit area and convoluted with the appropriate experimental resolution function.

Norm-conserving, nonlocal Dirac-Fock average relativistic effective pseudopotentials (AREP) \cite{trail2005_2,*trail2005} are used to approximate the ion cores.
Thus the obtained VMC APMDs include only the contribution from the valence electrons. We compute the core electron Doppler projections and their relative intensity as compared to total spectra with density functional theory (DFT) based methods. In the calculation of core spectra, we assume core orbitals of free atoms, and the shape of the positron orbital close to the nucleus is parametrized using linear-muffin-tin-orbital calculations \cite{MakkonenPRB2006}.  
The core spectra are calculated by using either the state-dependent model \cite{alatalo1996} or the state-independent model \cite{Daniuk1987,*Jarlborg1987}, and the LDA positron correlation potential and enhancement factor \cite{BoronskiPRB1986}. To estimate the total Doppler spectra with VMC we add the core electron projections to the QMC projection, with our final estimates being $\rho(p_L)=(1-\mu)\rho_{\text{val}}(p_L)+\mu \rho_{\text{core}}(p_L)$, where $\mu$ is the relative intensity of the core annihilation according to DFT. An alternative approach for estimating $\mu$ is to evaluate the valence and core annihilation rates using QMC and DFT, respectively (see Ref.~\cite{SimulaPRL2022}), but this gives only insignificant changes to the final results.  

\begin{figure*}
\includegraphics[scale=.551]{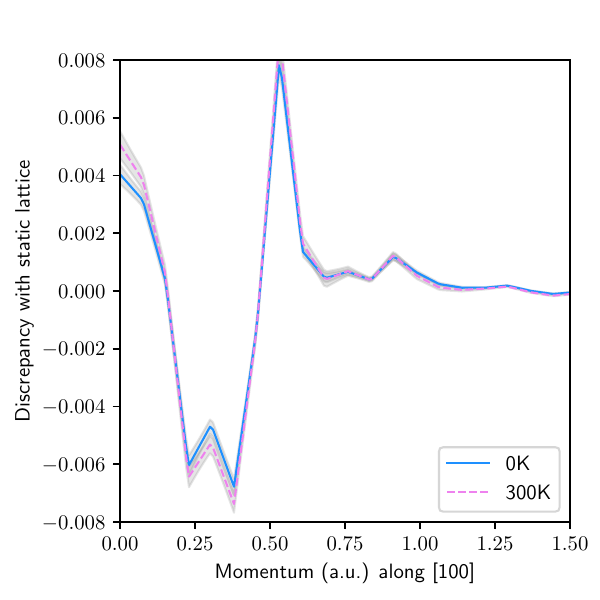}
\includegraphics[scale=.551]{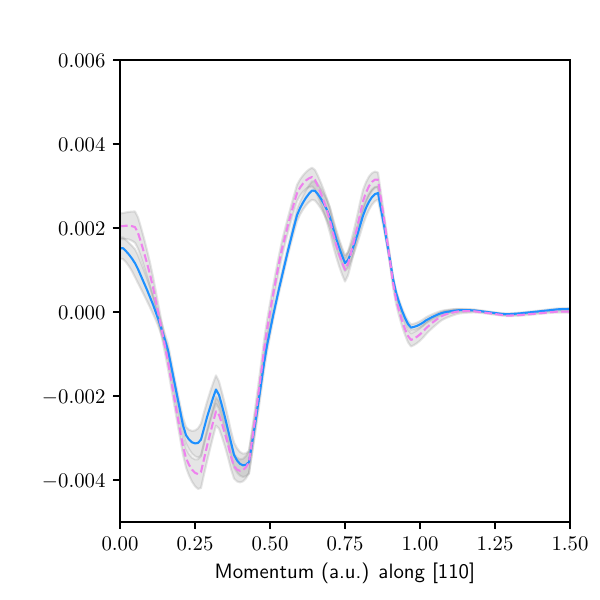}
\includegraphics[scale=.551]{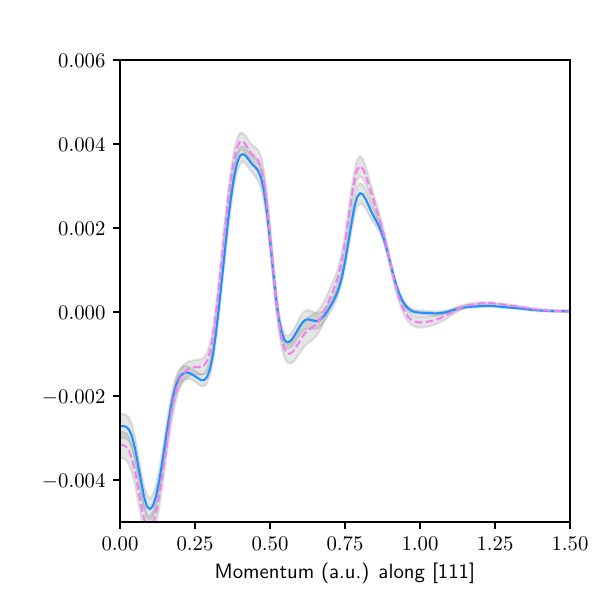}
\caption{\label{figure: vibrational data} Discrepancy of the calculated Doppler spectra of Si with vibrational corrections in $T=0$~K (blue) and T=$300$~K (violet) against the results from static lattice, calculated as $\rho_{\text{vib}}(p_L)-\rho_{\text{stat}}(p_L)$, where $\rho_{\text{stat}}(p_L)$ and $\rho_{\text{vib}}(p_L)$ are the Doppler spectra of the static and vibrating lattices, respectively. Results along the $[100]$ (left), $[110]$ (middle), and $[111]$ (right) lattice directions are shown.}
\end{figure*}

In addition to comparing our VMC results with experiments, we compare our results with more conventional DFT-based calculations and enhancement models for the APMD. We use the \textsc{vasp} code \cite{Kresse1996a,*Kresse1996b} and the projector augmented wave method \cite{Blochl1994,*Kresse1998} within the LDA  \cite{Perdew1981} for Si and GGA \cite{PerdewPRL1996} for C and AlN. We use the electron-positron LDA \cite{BoronskiPRB1986}, both state-dependent \cite{alatalo1996} and -independent models within the LDA \cite{Daniuk1987,*Jarlborg1987}, reconstructed PAW orbitals \cite{MakkonenJPCS2005,MakkonenPRB2006}, and we treat the core elecrons as described above.

We also use DFT to estimate the effect of lattice vibrations in Doppler spectra in bulk Si. In a $64$-atom simulation cell, we solve the phonon modes at the $\Gamma$ point and sample a set of atomic displacements with respect to the equilibrium structure, distributed according to the vibrational modes at temperatures of $0$ and $300$ K. Then we construct a set of simulation cells based on the sampled atomic displacements and estimate the APMD in each of the cells. This gives us APMD estimates distributed according to lattice vibrations at the two different temperatures. The method of sampling the atomic displacements is similar to the one used in Ref. \cite{SimulaPRL2022}, but here we used an algorithm implemented in \textsc{vasp}~\cite{KarsaiNJP2018}. It should be noted that this method does not include anharmonic effects.

\begin{figure*}%
  \includegraphics[scale=.8]{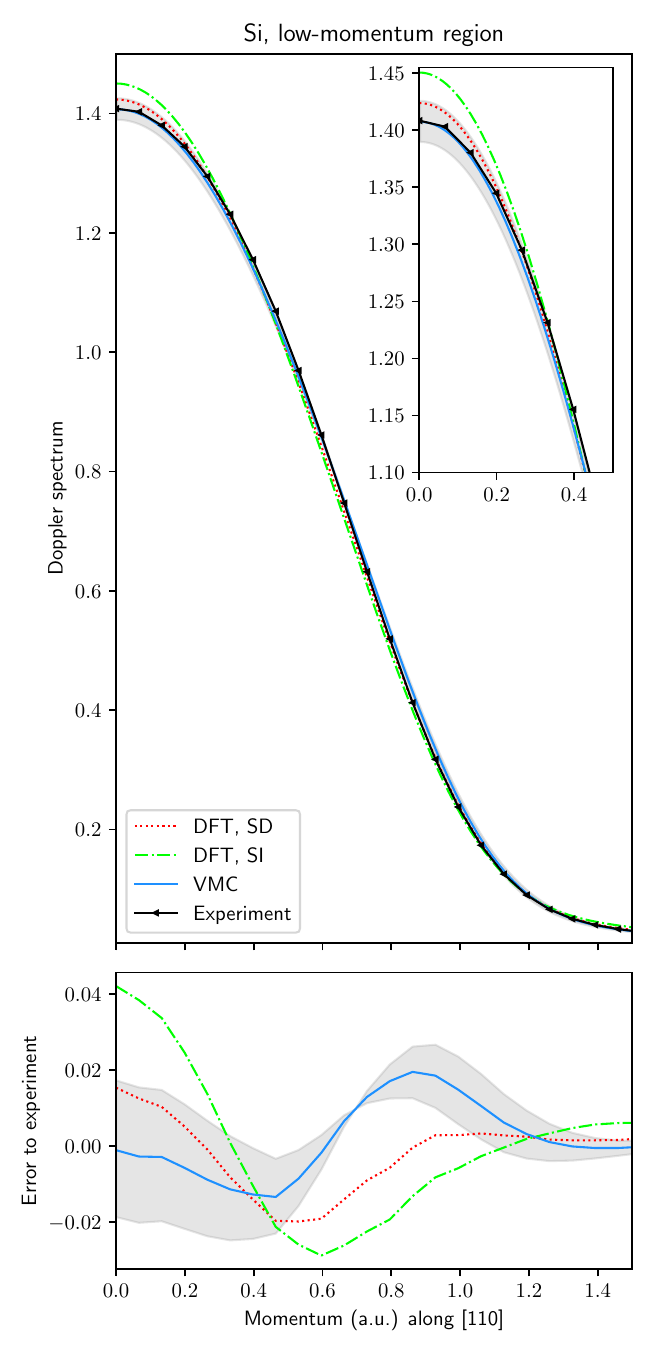}
  \includegraphics[scale=.8]{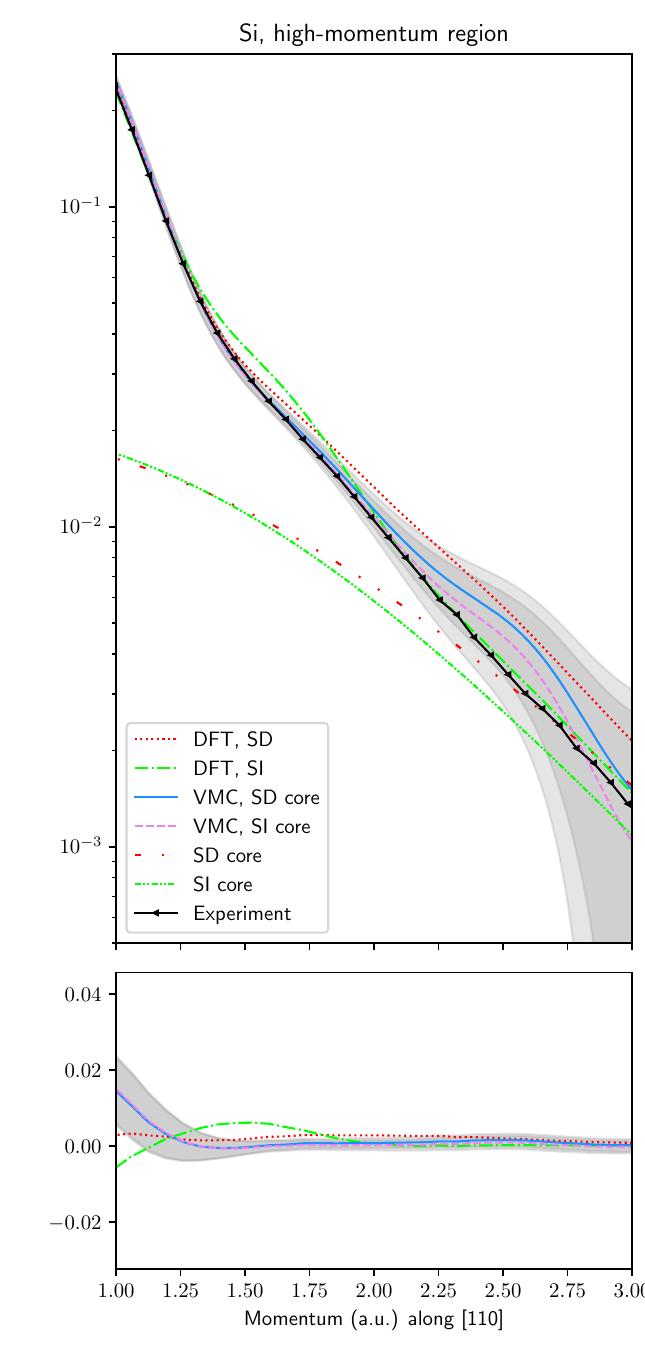}                                          
\caption{\label{figure:si_benchmark}  Low- and high-momentum parts (left and right panels, respectively) of the Si Doppler spectrum along the [110] direction. The data includes the experimental benchmark result~\cite{RummukainenPRL2005} (black), our VMC data (blue, and violet when the state-independent (SI) core spectrum is used), and for comparison DFT-based results calculated by the state-dependent (SD, red) and state-independent (pink) models. The core spectrum obtained from these models and used also in QMC is shown separately. In the left, the SD core spectrum was added to the QMC valence spectrum, in the right, QMC with both core spectra is shown.}
\end{figure*}

\section{Experimental}

We benchmark our simulations against experimental Doppler broadening data for Si and AlN samples that predominantly produce annihilations in the delocalized state in the lattice.

The reference Si measurement~\cite{RummukainenPRL2005} was performed using a slow positron beam and 2D coincidence Doppler setup with two HPGe detectors with a combined energy resolution of 0.92 keV full width at half maximum (FWHM). 
The measurement was performed on a (100) oriented Si substrate with $\mathbf{p}_L$ along the [110] crystal direction \footnote{The sample side orientations were confirmed by X-ray diffraction experiments for this paper, as they were incorrectly assumed to be [100] in Ref.~\cite{RummukainenPRL2005}.}.

The AlN reference measurement~\cite{MakiPRB2011} is also a 2D coincidence Doppler measurement performed using a slow positron beam, with a FWHM of 0.90~keV. The sample's growth direction is the c-axis direction [0001]. The precise alignment of the sample in the measurement is unknown and thereby we only can tell that the $\mathbf{p}_L$ direction is perpendicular to the c-axis, so we choose the direction of Doppler broadening spectrum as [$\bar{1}2\bar{1}0$]. The Doppler spectra of AlN are very isotropic in this plane and therefore the precise choice of the projection direction in simulations plays only a minor role.

\section{Results and discussion}

\subsection{Convergence with respect to system size and wave function}

Figure \ref{figure: c convergence} shows Doppler broadening results from VMC simulations in C and AlN projected in the $[100]$- and [$\bar{1}2\bar{1}0$]-directions, respectively, according to Eq.~(\ref{equation: doppler projection formula}). The results are obtained using SJ wave functions in $16$- and $54$-atom ($64$- and $216$-electron) simulation cells in C and $16$- and $72$-atom ($64$- and $288$-electron) simulation cells in AlN. Results from SJB simulations in $16$-atom simulation cells are also shown. Inspection of the different spectra in each of the materials show that the results obtained with different cell size and wave function are within each others' errorbars, and hence we can treat the SJ results in the smaller cells as being converged with respect to wave function form. Same conclusions were obtained from Doppler broadening calculations in Si. Discontinuities of the momentum density around the Fermi surface such as in Li \cite{yang2020} are not expected here, as instead of metals we are dealing with insulators and semiconductors. 

It should be noted that in AlN, the projection maximum in Fig.~\ref{figure: c convergence} is at non-zero momentum. This feature is qualitatively different from the monotonously degreasing shape DFT-based methods predict for AlN, although in Si a similar effect is visible at certain projection directions. We found this effect to take place in all projection directions in AlN with QMC APMD data. The effect can arise in QMC simulations due to explicit treatment of many-body correlations, but it is smeared out when the projection is convoluted with a resolution function of a detector used to accumulate reference experimental data, as can be seen in Fig.~\ref{figure:aln_benchmark}.

\subsection{The effect of lattice vibrations in Si}

Figure \ref{figure: vibrational data} shows the discrepancy of the Doppler spectra between renormalized, unconvoluted DFT results obtained with and without vibrational corrections  along the $[100]$, $[110]$, and $[111]$ directions in the Si lattice, at temperatures of $0$ and $300$~K. The discrepancies are calculated by removing the Doppler spectra of the static lattice from that of the vibrational calculation. It can be seen that there is not much effect due to vibrations on the Doppler spectra. The largest deviations can be seen around momentum of $0.5$ a.u.\ in the $[100]$-direction, where the discrepancy is still less than $1\%$ of the value of the normalized Doppler spectra of the static lattice. At momenta larger than $1$ a.u.\ the discrepancies settle very close to zero. The effect is the same at $0$ and $300$~K, indicating that the zero-point vibrations dominate also at room temperature. The discrepancy is also small when compared against the discrepancies between DFT and QMC, or between DFT and experiment, see below. 
Anharmonic effects at finite temperatures are not studied, but thermal expansion may have minor effects.

Figure \ref{figure: vibrational data} shows that the vibrations tend to broaden the Doppler spectra in the $[100]$ and $[110]$ directions, as the maximum negative discrepancy is seen around momentum values of $\sim 0.3$ a.u., while in $[111]$ direction this settles closer to zero, therefore tending to narrow the Doppler spectrum. However, as these effects are small as mentioned above.

\subsection{Benchmark against experiments and model calculations}

We now compare our results against the 2D coincidence beam measurements for Si~\cite{RummukainenPRL2005} and AlN~\cite{MakiPRB2011}. These experimental references are the best currently available given the computational constraints, and further data for e.g.\ diamond are too noisy to allow us to draw conclusions on whether DFT or QMC methods provide better results.

\subsubsection{Si}

Figure~\ref{figure:si_benchmark} shows a comparison of the low- and high-momentum regions of the Doppler spectrum between VMC and DFT models and the experimental reference~\cite{RummukainenPRL2005}. The low- and high-momentum regions are determined mainly by valence and core electrons, respectively. The high-momentum region contains also information relevant for identifying atom types around annihilation sites.

At low momenta we find that the VMC result agrees with experiment within statistic error bars. At higher momenta the agreement remains good, but at momenta higher than $p=2.5$~a.u the core spectra, calculated with DFT, start to dominate the results and the statistical error of the VMC results starts to be large when compared against the projection values.
The choice of the correlation model used for the core electrons of Si does not play a role in the momentum range where we can reliably compare experiment and theory ($p<2.5$~a.u.).

\begin{figure*}%
\includegraphics[scale=.8]{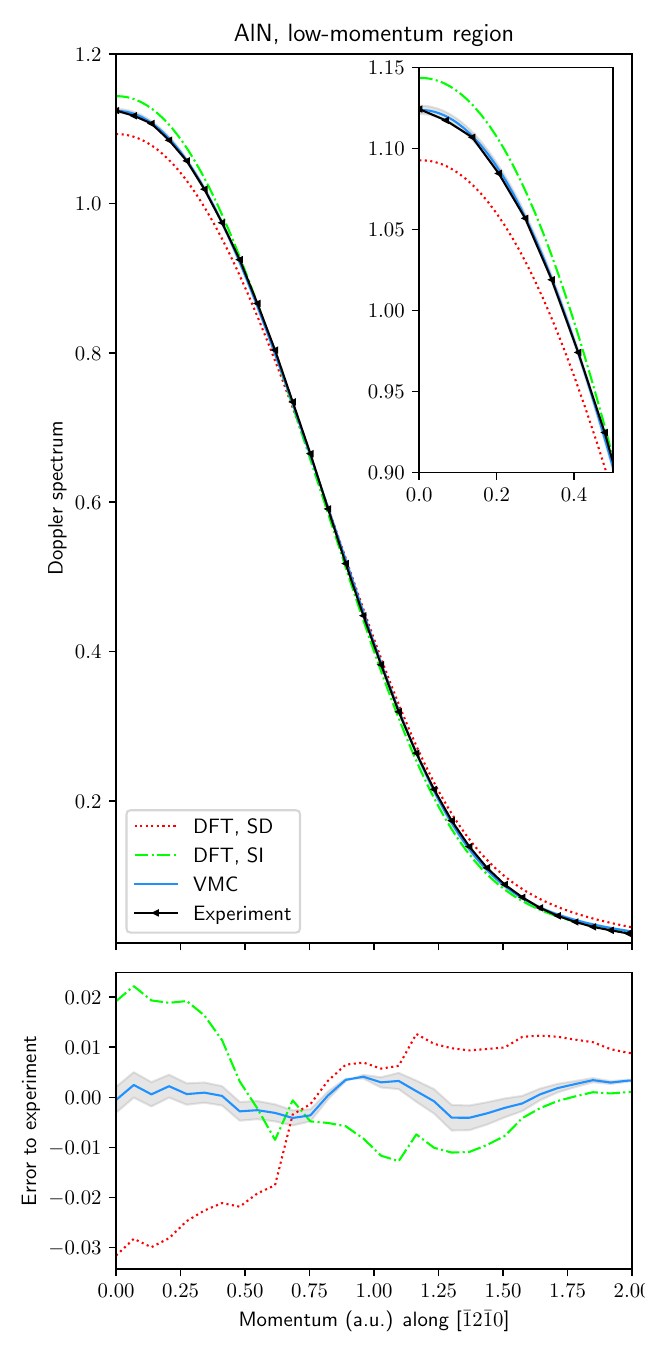}
\includegraphics[scale=.8]{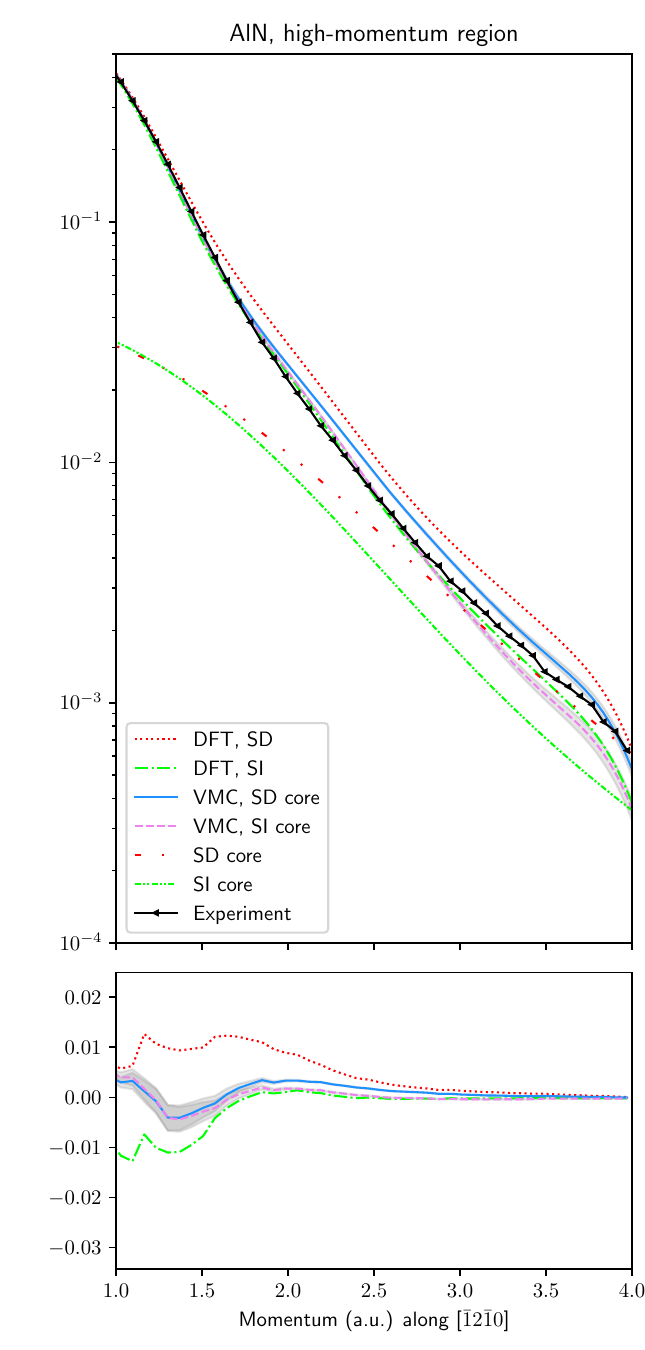}
\caption{\label{figure:aln_benchmark} Low- and high-momentum parts (left and right panels, respectively) of the AlN Doppler spectrum along the [$\bar{1}2\bar{1}0$] direction. The data includes the experimental benchmark result~\cite{MakiPRB2011} (black), our VMC data (blue, and violet when the state-independent (SI) core spectrum is used), and for comparison the DFT-based results calculated by the state-dependent (SD, red) and state-independent (pink) models. The core spectrum obtained from these models and used also in QMC is shown separately. In the left, the state-dependent core spectrum was added to the QMC valence spectrum, in the right, QMC with both core spectra is shown.}
\end{figure*}

Figure~\ref{figure:si_benchmark} shows that the agreement of the DFT-based models (state-dependent and -independent, both applied with Boro\'nski-Nieminen LDA \cite{BoronskiPRB1986}) with experiment is worse than that of VMC, and the results do not lie within statistical error bars. The list of models compared with here is not exhaustive, but the point in this comparison is that the spread between these two results gives an idea of the typical error due to the choice of the correlation model and functional used in the calculation. We want to stress that, concerning the enhancement factor parametrizations in the literature, only the various LDA parametrizations (see, for example. Refs.~\cite{BoronskiPRB1986} and \cite{DrummondPRL2011}) and a single generalized gradient model~\cite{BarbielliniPRL2015} are genuinely parameter-free. 

At low momenta the state-dependent model agrees better with experiment and VMC, whereas at higher momenta its well-known overestimation of intensity~\cite{MakkonenPRB2006} when applied within the LDA is apparent and the state-independent model provides a better agreement with the benchmark.

\subsubsection{AlN}

A similar comparison as for Si is shown for AlN in Fig.~\ref{figure:aln_benchmark}. The VMC result displays excellent agreement with the experimental reference~\cite{MakiPRB2011}, whereas the DFT-based model calculations deviate from the experiment and VMC results. At high momenta the model used for the core plays a larger role here, as the deviation between the core spectra of the state-dependent and -independent models is larger than in Si. The state-independent model gives a better agreement with experiment than the state-dependent model.

Concerning the DFT-based models, at low momenta both state-dependent and state-independent models perform equally well with opposite errors. At high momenta we again see the typical overestimation of the Doppler spectrum by the state-dependent model and the LDA.

\section{Conclusions}

In this work, a VMC method for computing annihilating-pair momentum densitites in insulator and semiconductor materials was presented. This method can be applied in metals, but this requires a more accurate sampling of the momentum grid, meaning that one should have a larger simulation cell or a denser twist grid. The VMC results were compared against state-dependent and state-independent models in the framework of DFT and against experimental Doppler broadening measurements.

The convergence with respect to simulation cell size was tested. It was found that backflow is not necessary for obtaining accurate and convergent results. We also tested the effects of lattice vibrations on Doppler spectra, and found that it is sufficient to consider a static lattice with equilibrium atomic structure. These tests show that it is possible to perform accurate VMC calculations of Doppler spectra with relatively low computational costs.

The VMC results match better with experimental Doppler broadening spectra than with results from previous DFT-based methods. This shows how explicit inclusion of the electron-electron and electron-positron correlation effects can be used to improve the correspondence of experiment and theory in positron-based spectroscopies. Together with our previous QMC results for positron lifetimes in solids \cite{SimulaPRL2022}, this article provides a consistent picture concerning the accuracy and shows how QMC can be used as a practical tool to support positron spectroscopies when high accuracy is needed. 

A recent study \cite{simula2023nv} shows that the same QMC method that is capable of predicting positron lifetimes and Doppler broadening spectra accurately can be utilized in studying the excited states of the nitrogen-vacancy center in diamond. Hence the simultaneous QMC description of the electronic structure and the trapped positron state should be tested. There are also many systems with less static character in correlations, which can be problematic to treat with DMC, than in the nitrogen-vacancy center, where positron states can be immediately described with single-determinant wave functions and methods presented here and in Ref.~\cite{SimulaPRL2022}. In the near future we expect QMC to be applicable to simulations of positrons in both a delocalized bulk state and in trapped states at open-volume defects. This, among other promising routes such as simulation of positrons at surfaces or interfaces, shows the potential for very interesting applications of QMC in support of positron spectroscopies.

\begin{acknowledgments}
We acknowledge the generous computational resources provided by CSC (Finnish IT Centre for Science), in particular to Mahti and Lumi clusters.  This work was partially supported by the Academy of Finland Grants No.\ 285809, No.\ 293932, No.\ 319178, No.\ 334706, and No.\ 334707.           
\end{acknowledgments}    



%

\end{document}